\documentclass{aa}
\usepackage{txfonts}
\usepackage{graphicx}
\usepackage{longtable}
\usepackage{latexsym}
\usepackage{amssymb}
\usepackage{lscape}
\input{psfig.tex}

\begin{document}
\title{The VVDS data reduction pipeline: introducing VIPGI, 
the VIMOS Interactive Pipeline and Graphical Interface}

\author{M. Scodeggio \inst{1}
\and P. Franzetti \inst{1}
\and B. Garilli \inst{1}
\and A. Zanichelli \inst{2}
\and S. Paltani \inst{3}
\and D. Maccagni\inst{1}
\and D. Bottini \inst{1}
\and V. Le Brun \inst{3}
\and T. Contini \inst{4}
\and R. Scaramella \inst{2}
\and C. Adami \inst{3}
\and S. Bardelli \inst{5}
\and E. Zucca \inst{5}
\and L. Tresse \inst{3}
\and O. Ilbert \inst{3}
\and S. Foucaud \inst{1}
\and A. Iovino \inst{6}
\and R. Merighi \inst{5}
\and G. Zamorani \inst{5}
\and I. Gavignaud \inst{4,12}
\and D. Rizzo \inst{4}
\and H.J. McCracken \inst{7,8}
\and O. Le F{\`e}vre\inst{3}
\and J.P. Picat \inst{4}
\and G. Vettolani \inst{2}
\and M. Arnaboldi \inst{9}
\and S. Arnouts \inst{3}
\and M. Bolzonella \inst{10}
\and A. Cappi \inst{5}
\and S. Charlot \inst{7,11}
\and L. Guzzo \inst{6}
\and B. Marano \inst{10}
\and C. Marinoni \inst{3}
\and G. Mathez \inst{4}
\and A. Mazure \inst{3}
\and B. Meneux \inst{3}
\and R. Pell{\`o} \inst{4}
\and A. Pollo \inst{6}
\and L. Pozzetti \inst{5}
\and M. Radovich \inst{9}
}

\institute{IASF-INAF - via Bassini 15, I-20133, Milano, Italy
\and
IRA-INAF - Via Gobetti,101, I-40129, Bologna, Italy
\and
Laboratoire d'Astropysique de Marseile, UMR 6110 CNRS-Universit\'e de
Provence,  BP8, 13376 Marseille Cedex 12, France
\and
Laboratoire d'Astrophysique de l'Observatoire Midi-Pyr\'en\'ees (UMR
5572) - 14, avenue E. Belin, F31400 Toulouse, France
\and
INAF-Osservatorio Astronomico di Bologna - Via Ranzani,1, I-40127, Bologna, Italy
\and
INAF-Osservatorio Astronomico di Brera - Via Brera 28, Milan, Italy
\and
Institut d'Astrophysique de Paris, UMR 7095, 98 bis Bvd Arago, 75014
Paris, France
\and
Observatoire de Paris, LERMA, 61 Avenue de l'Observatoire, 75014 Paris,
France
\and
INAF-Osservatorio Astronomico di Capodimonte - Via Moiariello 16, I-80131, Napoli,
Italy
\and
Universit\`a di Bologna, Dipartimento di Astronomia - Via Ranzani,1,
I-40127, Bologna, Italy
\and
Max Planck Institut fur Astrophysik, 85741, Garching, Germany
\and
European Southern Observatory, Karl-Schwarzschild-Strasse 2, D-85748
Garching bei München, Germany}
\date{}

\abstract{The VIMOS VLT Deep Survey (VVDS), designed to measure
150,000 galaxy redshifts, requires a dedicated data reduction and
analysis pipeline to process in a timely fashion the large amount of
spectroscopic data being produced. This requirement has lead to the
development of the VIMOS Interactive Pipeline and Graphical Interface
(VIPGI), a new software package designed to simplify to a very high degree
the task of reducing astronomical data obtained with VIMOS, the
imaging spectrograph built by the VIRMOS Consortium for the European
Southern Observatory, and mounted on Unit 3 (Melipal) of the Very
Large Telescope (VLT) at Paranal Observatory (Chile). VIPGI provides
the astronomer with specially designed VIMOS data reduction functions,
a VIMOS-centric data organizer, and dedicated data browsing and
plotting tools, that can be used to verify the quality and accuracy of
the various stages of the data reduction process. The quality and
accuracy of the data reduction pipeline are comparable to those
obtained using well known IRAF tasks, but the speed of the data
reduction process is significantly increased, thanks to the large set of
dedicated features.  In this paper we discuss the details of the MOS
data reduction pipeline implemented in VIPGI, as applied to the
reduction of some 20,000 VVDS spectra, assessing quantitatively the
accuracy of the various reduction steps. We also provide a more
general overview of VIPGI capabilities, a tool that can be used for
the reduction of any kind of VIMOS data. 
}

\offprints{M. Scodeggio}
\authorrunning{Scodeggio et al.}
\titlerunning{the VVDS data reduction pipeline}

\maketitle

\section{Introduction}
\label{sec:intro}

Over the last few years the number of large telescopes available to
the astronomical community has rapidly increased, together with the
multiplexing capabilities of their instruments.  While a normal
long-slit spectrograph on a 4-meter class telescope could produce a
few tens of spectra per night of observation, today a spectrograph
like VIMOS at the VLT, or 2dF at the AAT, can obtain several thousands
of spectra per night. This productivity increase has rendered obsolete
traditional methods of data reduction and analysis, at least as long
as these data must be reduced and analyzed in a timely fashion.  It is
clearly necessary to automatize as much as possible these operations,
to increase the speed with which they can be carried out, but without
sacrificing the capability of analyzing in detail the results of the
various operations, and eventually manually intervene to change the
way some of these operations are carried out. Moreover it is necessary
to develop some efficient and rigorous data organizer and archiver, so
that the available files and data would not be lost among hundreds or
thousands of similar data and files.  General-purpose astronomical
software packages are not well equipped for these tasks, as witnessed
by the ad-hoc data reduction pipelines developed for most large
spectroscopic surveys, both in the case of multi-slit (see for example
Colless et al. 1990 for the LDSS; Le F{\` e}vre et al. 1995 for the
CFRS) and multi-fiber observations (see for example Colless et
al. 2001 for the 2dF; Stoughton et al. 2002 for the SDSS).

In particular, multi-fiber observations based surveys in the recent
past have been very successful in processing large amount of
spectroscopic data, but to their advantage was the fact that they were
observing relatively bright galaxies at low redshift. Therefore the
observations were carried out in the blue part of the optical
spectrum, a wavelength domain contaminated by only a few strong sky
emission lines. And the relative brightness of the observed galaxies
makes the sky-subtraction process a non critical one, whereby the
determination of the sky spectrum required only a few percent accuracy
level. Significantly more challenging is the data processing task for
a survey extending to higher redshift than that covered by the 2dF
(Colless et al. 2001) and SDSS (Strauss et al. 2002) surveys. In this
case it is necessary to derive spectra of very faint galaxies, with a
surface brightness only a few percent of the sky one, and in the red
part of the optical spectrum, a wavelength domain contaminated by many
strong sky emission features. Early work in this challenging domain
was provided by the CFRS team (Le F{\` e}vre et al. 1995), but
nowadays the high data throughput of 8-meter class telescopes require
significant research and development efforts to obtain a spectroscopic
data reduction pipeline capable of properly handling the data these
telescopes are obtaining.

Among the last-generation spectrographs VIMOS is perhaps the most
challenging in terms of data production, being currently the
instrument with the highest multiplexing capabilites available to any
astronomer. VIMOS is an imaging spectrograph mounted on the Unit 3
telescope (Melipal) of the Very Large Telescope (VLT) at the Paranal
Observatory, in Chile (see Le F{\` e}vre et al., 2000; 2002 for a
detailed description of the instrument and its capabilities). It has
been specially designed to be a survey instrument, and therefore its
multiplexing capabilities have been pushed to the maximum: its field
of view, divided in four separate quadrants, covers almost entirely
the unvignetted area of the Nasmyth VLT focal plane, and during a
single exposure, up to 1000 spectra can be obtained in MOS mode (6400
in IFU mode).  Difficulties for the user in the data reduction process
already begin when trying to find one's way among the large number of
raw science and calibration files, and increase going further along
the reduction process, due to the number of calibrations and
corrections that have to be applied to the data.  To give an example,
even just a single pointing in MOS mode, with its minimal set of
calibration data (five bias, five flat field and one arc lamp
exposure), is equivalent to 48 FITS files being produced (raw data
only).

For this reason we were contracted by the European Southern
Observatory (ESO) to deliver to them all the elements necessary to
build a VIMOS-specific automatic data reduction pipeline. In
collaboration with the ESO Data Management Division, we have therefore
developed a C library of data reduction procedures, the VIMOS Data
Reduction Software (DRS, see Scodeggio et al. 2000), which is now
being used for the on-line reduction of VIMOS data at ESO. Because of
its on-line usage, this ESO pipeline must work in a completely
automatic fashion, but while this is an optimal choice to obtain a
quick quality assessment for VIMOS observations, it is a far less
acceptable one for a careful and complete science data reduction
pipeline. Faced with the task of reducing some 100,000 spectra
observed as part of the VLT VIMOS Deep Survey (VVDS, see LeF{\` e}vre
et al. 2004 for a detailed description of the survey), we have used
the DRS to produce a semi-automatic data reduction pipeline, the VIMOS
Interactive Pipeline and Graphical Interface (VIPGI). With VIPGI we
have kept the DRS capability for a very fast data reduction process,
but we also have added many data reduction quality control points, a
user friendly graphical interface, and a simple but effective method
to organize the raw and reduced data.

VIPGI was designed specifically to carry out the reduction of the VVDS
data, and its performance while performing this task is the focus of
this paper. But VIPGI capabilities are far more general. It can be
used to reduce VIMOS data of any kind, and not only MOS data. In
particular, it can be used to reduce VIMOS IFU data, a task far more
complex than the relatively simpler reduction of MOS data, and which
is described in detail in a separate paper (Zanichelli et al. 2004).
But VIPGI capabilities are general enough, at least as far as MOS data
are involved, that it could be used as a generic pipeline tool for the
reduction of MOS data from any spectrograph, and work is already under
way to develop this functionality (see Paioro et al. 2004).  So far
VIPGI has been extensively tested while reducing more than 20,000 MOS
spectra for the VVDS, and a smaller number of MOS and IFU spectra from
other, independent projects. At the moment it is not publicly
released, but it can be used either in Milano (Italy) or Marseille
(France) by members of the ESO community under the supervision of a
VIRMOS Consortium astronomer, who can provide guidance on how to
handle non standard situations\footnote{see
http://cosmos.mi.iasf.cnr.it/marcos/vipgi/vipgi.html}.

In this paper we discuss some of VIPGI general concepts
(sections~\ref{sec:VIPGI} and \ref{sec:calibs}), and then focus on the
reduction of VVDS MOS data, and the quality of the final VVDS spectra
(sections~\ref{sec:reduction} and \ref{sec:quality}). The data
organizer and the data browsing and plotting facilities are briefly
described in sections~\ref{sec:organizer} and \ref{sec:plotting}.

\section{The VIMOS Interactive Pipeline and Graphical Interface}
\label{sec:VIPGI}

The core of the VIPGI pipeline is composed of a set of routines
performing the data reduction, coded using the C language to obtain
the maximum speed for this computationally intensive task.  Routines
range from the basic ''opening a file and reading its content'' to
wavelength calibration, spectral extraction and IFU 3D data cube
reconstruction, and are organized in a relatively small number of
reduction recipes. Generic tasks like the handling of FITS files, or
of the World Coordinate System and the detection of stars within
imaging exposures (for photometric or astrometric calibration
purposes) are devoted to special purpose external software packages
(the CFITSIO and WCSTools libraries and SExtractor, respectively) that
have been included within VIPGI.

When designing recipes, we have tried to group together steps which
are normally always executed in the same sequence: for example, bias
subtraction, flat field correction and bad pixels cleaning have been
grouped into one recipe.  However there is no single "do it all"
recipe that can be fed with a bunch of raw data frames to produce
completely reduced images or spectra, as we assume that the
astronomers will need and want to check at least some of the
intermediate data reduction steps. To help astronomers keep the
details of the data reduction process under control, the detailed
behavior of each recipe can be customized via a set of input
parameters, that are stored in a parameter file.

All recipes have been written to work with files in FITS format. To
avoid increasing the already large number of files, the different
reduction mid-products together with the various calibration tables
needed for the reduction process are ``appended'' as extensions to the
original FITS file, instead of creating independent files. As a
consequence, the results of the reduction of one spectroscopic
exposure are contained within 4 files only (one per quadrant), each
file containing up to 12 FITS image and binary table extensions. Thus
these files can easily be over 100 MB in size.

The C recipes automatize to a very large extent the task of reducing
VIMOS data but they do not address at all two important and
problematic areas of the global data reduction activity. The first one
is the need of organizing the large volume of VIMOS data and the
second one is the need for a quick and easy browsing of the data at
the various stages of data reduction.  It is mainly to address these
two needs that we have designed the VIPGI graphical user interface.
It provides a VIMOS-specific data organizer, designed to help the user
to select the correct data to use at all stages of the data reduction
process, a simple interface to the underlying C reduction procedures,
tools that allow a simple and quick browsing trough the VIMOS data and
powerful plotting tools to view and analyze 1D and 2D extracted
spectra.

The graphical user interface has been coded using the Python language,
and its standard Tkinter graphical interface to the Tk set of widgets.
All plotting functions use the BLT library for the actual drawing of
the plot elements. No built-in images display tool is provided, but
users can configure VIPGI to use their preferred one. VIPGI features
are presented in the following sections. In the web site
{http://cosmos.mi.iasf.cnr.it/marcos/vipgi/vipgi.html} additional
informations and screenshots can be found to better understand all
VIPGI features. 

In total the software that was developed specifically for this project
is composed of some 151,000 lines of C code, and of 16,000 lines of Python
code. The overall performance of the VIPGI pipeline is of course
dependent on the computer hardware being used, and on the kind of data
being reduced. As an example, we quantify the time requirements for
the reduction of a typical VVDS deep pointing (4 masks for a total of some
550 slits, 10 jittered exposures to be combined), on a Linux PC
equipped with an AMD Athlon XP 2800+ CPU and 1 GB of RAM.
It takes approximately 25 CPU seconds to obtain the complete wavelength
calibration for the 4 quadrants, and 1170 CPU seconds to complete the
reduction of the science data, again for the 4 VIMOS quadrants. From
the end user point of view, it takes approximately 1 hour to go from
the raw data to a set of fully reduced spectra,  including the CPU
time required to run the reduction recipes, the time spent
selecting the appropriate input files for those recipes, and the
visual inspection and verification of the instrument calibrations
described here below.

\section{The instrument calibrations}
\label{sec:calibs}

VIPGI recipes require an already existing calibration of the
instrument properties as part of their input. Therefore a fundamental
component of whole pipeline architecture is the VIMOS instrument
model, which analytically describes the main calibration relations
required for the extraction of object spectra from VIMOS spectroscopic
observations. The model is separated into three different
components:\\
1) the Optical Distortion Model, which provides a mapping between
positions on the VIMOS focal plane and pixel coordinates on the CCD
frame. This mapping is obtained for an arbitrarily fixed wavelength,
and is described by two independent polynomial relations. Since VIMOS
spectra are dispersed along the CCD columns, the relation that gives
the x pixel coordinate as a function of position in the focal plane is
wavelength independent, and is influenced only by the optical and
mechanical layout of the instrument. The relation that gives the y
pixel coordinate is instead dependent also on the choice of the
reference wavelength.\\
2) the Curvature Model, which provides a description of the
geometrical shape of each spectrum on the CCD, to allow for its
tracing and extraction. This model is obtained using a low order
$(\leq 2)$ polynomial.\\
3) the Inverse Dispersion Solution, which provides the mapping between
wavelength and pixel coordinates along the geometrical shape traced by
the Curvature Model, and measured as offsets with respect to the
reference wavelength and pixel position defined by the Optical
Distortion Model. Also this mapping is described by a relatively low
order polynomial, with the order depending on the grism used to obtain
the data, and therefore on the observations spectral resolution.

The instrument model can be used at two different levels of detail to
obtain a description of VIMOS data. The local model provides the
highest level of detail, as it describes each MOS slit or IFU fiber
spectrum individually. The global model instead provides a global
description for one whole CCD frame, which is significantly more
robust than that provided by the local model, as it is derived using
many tens (MOS case) or hundreds (IFU case) or spectra, although at a
reduced level of detail. The Optical Distortion Model, by its very
nature, exists only at the global level.

The starting point for each recipe is always a global model, whose
parameters are stored in the raw data FITS file header when the
observations are carried out. This model is used to derive a local
model for each spectrum, which is then refined fitting its parameters
to the real data. Finally, an updated global model is obtained fitting
the parameters of all available local models. In this way the best
possible calibration is obtained for each individual VIMOS
exposure. Still, with accurately calibrated global and local models,
it is important to decide which one to use for the reduction of
scientific observations. The choice actually used within VIPGI is
described in detail in the following sections.

Since the four VIMOS quadrants correspond to four physically distinct
cameras within the instrument, each quadrant is characterized by its
own instrument model. Therefore all calibration and science data
reduction procedures are carried out on each quadrant data
independently, and only at the end of the data reduction process, and
only in some special cases (like IFU data cube reconstruction) the
data from all four quadrants are brought together into a single
data-set.

\begin{figure}
\resizebox{\hsize}{!}{\includegraphics{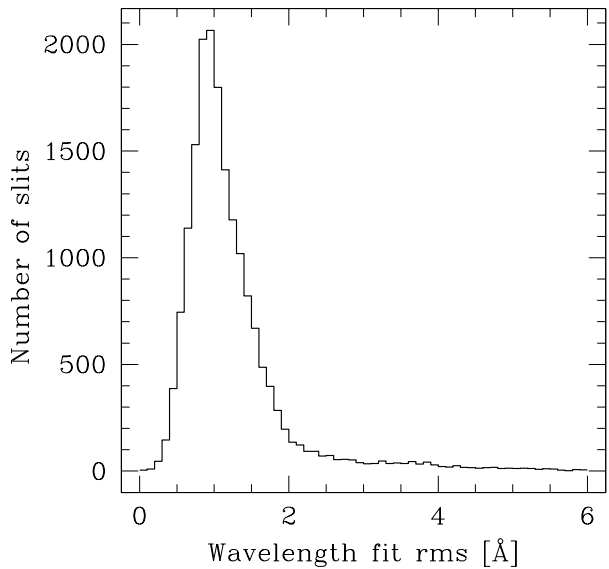}}
\caption{The distribution of wavelength calibration rms residuals for
  individual MOS slits. All 16,936 slits reduced thus far are included
  in the plot. The median residual rms is 1.005 \AA.}
\label{fig:lambdacal_rms}
\end{figure}

\subsection{The location of spectra on the CCD frame}
\label{sec:spectralocation}

The Optical Distortion and Curvature models are derived from the
accurate determination of the spectra location on the CCD frame. Using
MOS slit positions that are stored in the raw data FITS file header
(or IFU fiber end positions which, being fixed, are stored in a
permanent calibration table), and the global instrument model
parameters a first guess for the position of each slit or fiber
spectrum on the CCD is obtained. Starting from this position a search
is made for the edges of the illuminated area created by the spectrum
on the CCD. The shape of these edges is fitted with a polynomial, to
provide the updated local determination of the Curvature Model, and a
very precise determination of the spectrum position on the CCD. Then
curvature model parameters and position measurements for all slits are
fitted with polynomials to provide the updated global determination
for the Optical Distortion and Curvature Model.

MOS spectra are generally located on the CCD with a typical
uncertainty of 0.2 pixels, provided that the mask layout derived
automatically by the VIMOS Mask Preparation Software (VMMPS, see
Bottini et al. 2004) is not altered by the astronomer by adding
manually slits on the mask. VMMPS is in fact enforcing the presence of
a minimum gap between adjacent MOS slits, in order to make the
measurement of each spectrum edges possible.  IFU fiber spectra are
not so cleanly separated, as there are no sharp slit edges to
locate. In this case the location is carried out identifying the peaks
of the light distribution, and the tracing is then carried out
following the minimum of the light intensity between two adjacent
peaks, with a typical uncertainty of half a pixel (see Zanichelli et
al. 2004 for details).

With the exception of IFU data taken with high resolution grisms, the
accuracy with which the instrument global and local model can describe
the location of spectra on the CCD is comparable, and therefore by
default VIPGI uses the more robust global model to carry out this
description. Using VIPGI data browsing facilities it is possible to
carry out an accurate visual check of the quality of spectra location,
by displaying a raw VIMOS data frame with superposed the reconstructed
location of each spectrum.

\subsection{The wavelength calibration}
\label{sec:lambdacal}

The Inverse Dispersion Solution, which provides the wavelength
calibration for VIMOS spectra, is derived by measuring the position on
the CCD of a number of He, Ne or Ar lamp lines. The list of lines to
be measured is provided by a pre-defined Line Catalog, while the
location of the lamp spectra is described by the previously derived
Instrument Model, using the local model obtained for each slit or
fiber spectrum. The two-dimensional spectrum produced by a MOS slit is
separated into n one-dimensional spectra, one for each pixel spanned
by the length of the slit (for IFU fiber spectra only the central
pixel spectrum is extracted). Within these spectra the position of
lamp lines is measured, computing their baricenter within a
pre-defined extraction window, and measured positions are fitted
against the known line wavelengths using a low order polynomial
function and an iterative sigma clipping procedure. This fit provides
the coefficients for the local model Inverse Dispersion Solution for
each slit. 

\begin{figure}
\resizebox{\hsize}{!}{\includegraphics{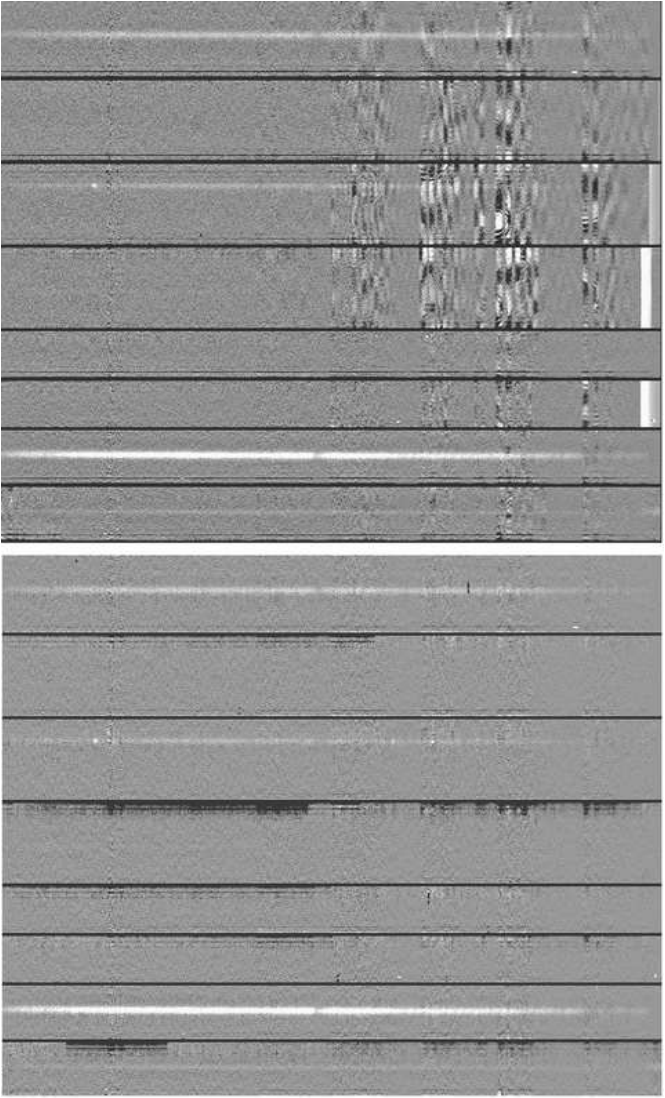}}
\caption{Example of the efficiency of the fringing subtraction
  procedure. The top half of the figure shows the result of combining
  10 jittered VVDS exposures without the subtraction of the fringing
  residuals. The bottom half shows the result of combining the same 10
  exposures with the subtraction. Only a small portion of the entire
  CCD frame is shown.
}
\label{fig:fringing}
\end{figure}

The accuracy of the wavelength calibration changes slightly from grism
to grism, but the {\it rms} residuals around the best fitting relation
typically amount to one fifth of a pixel. For the VVDS data, obtained
with the Low Resolution Red VIMOS grism, which produces spectra with a
linear dispersion of 7.14 \AA/pixel, a fit with a third degree
polynomial results in a wavelength calibration with a median {\it rms}
residual of 1.005 \AA. Figure~\ref{fig:lambdacal_rms}  shows the
distribution of rms residuals for a total of 16,936 slits: although
the distribution shows the presence of a tail extending to high rms
residual values, only approximately 0.9 percent of the slits show a
wavelength calibration with rms residuals larger than 3.0 \AA.
Contrary to the case of spectra location, for wavelength calibration
the local instrument model always provides the most accurate
calibration information, and therefore it is the one used in the
reduction of scientific observations. Using VIPGI data browsing and
plotting tools it is possible to carry out an accurate visual check of
the quality of the wavelength calibration, and also re-compute such
calibration for individual slits or fibers, after having eliminated
from the list of spectral lines significantly deviant ones.

The accuracy and stability of the wavelength calibration are obviously
of the greatest importance for a spectroscopic survey like the
VVDS. However the accuracy with which the Inverse Dispersion Solution
maps the true wavelength calibration of VIMOS observations cannot be
quantified solely on the basis of the rms residual values obtained in
the calibration procedure. In fact changes in temperature or flexures
within the instrument might significantly alter this mapping between
the times when the science and the calibration exposures are taken
(calibration lamp exposures are typically taken only at the end of a
sequence of exposures, and sometimes during the following day).
Therefore it is important to re-consider the issue of the wavelength
calibration accuracy on the basis of the reduced science data and
redshift measurements, and we do that in Section~\ref{sec:quality}.

\section{Science data reduction}
\label{sec:reduction}

The first step in the reduction of VIMOS science data is the canonical
preliminary reduction of CCD frames, which includes prescan level and
average bias frame subtraction, trimming of the frame to eliminate pre
and overscan areas, interpolation to remove bad CCD pixels, and flat
fielding.  For VVDS data, taken with the Low Resolution Red VIMOS
grism, the flat fielding of the data is not carried out, since flat
field exposures, obtained using an internal halogen lamp which
illuminates the reflective cover of the telescope Nasmyth focus
shutter screen, show the presence of a significant amount of
fringing. The spatial frequency of this fringing pattern is too high
to be reliably removed with some surface fitting procedure, and the
pattern present in flat field exposures results completely different
from the one present in science exposures, due to the different
spectral energy distribution of the halogen lamp and of the night
sky. As the noise introduced by flat fielding VVDS data with such
fringing-rich flat field frames is actually larger and more structured
than that produced by the CCD pixel to pixel sensitivity variations,
it was decided to eliminate the flat fielding altogether from the VVDS
data reduction scheme.

After the preliminary reduction step, subsequent data reduction steps
are carried out on all MOS slits individually, one slit at a time.
Since VIMOS is known to suffer from the presence of some flexures
within the instrument, the first among such steps is a refinement of
the wavelength calibration. This operation is carried out using a
number of skylines, by comparing their known wavelength with the one
derived using the local model Inversion Dispersion Solution for the
given slit. If any discrepancy is measured, an offset is introduced
into the Inverse Dispersion Solution to compensate it. In practice
this is a virtual offset to be applied to all pixel coordinates of a
given slit or fiber spectrum at the moment of its extraction,
discussed here below.

The following steps in the data reduction procedure include object
detection and sky subtraction from each MOS slit spectrum. The object
detection procedure is based purely on the data themselves, and does
not use any prior information from the MOS mask design procedure to
locate object spectra within the data. The raw data slit spectrum is
collapsed over a user-defined wavelength interval, following the
geometrical shape defined by the Curvature Model for the given slit,
to produce a slit cross-dispersion profile. A robust determination of
the average signal level and its {\it rms} variations in this profile
is obtained using an iterative sigma-clipping procedure, and objects
are detected as groups of contiguous pixels all above a given
detection threshold, expressed in units of the signal level {\it rms}
variations. For VVDS data we use a detection threshold of $2\sigma$,
and a minimum object size of 3 pixels. Possible spurious detections,
often taking place close to the slit edges, are removed at a later
stage, during the visual inspection of the extracted spectra. All
sections of the slit cross-dispersion profile devoid of objects are
considered sky regions, and are used in the original raw data slit
spectrum to derive a median estimate of the sky spectrum, which is
then subtracted from the whole slit data.

\begin{figure}
\resizebox{\hsize}{!}{\includegraphics{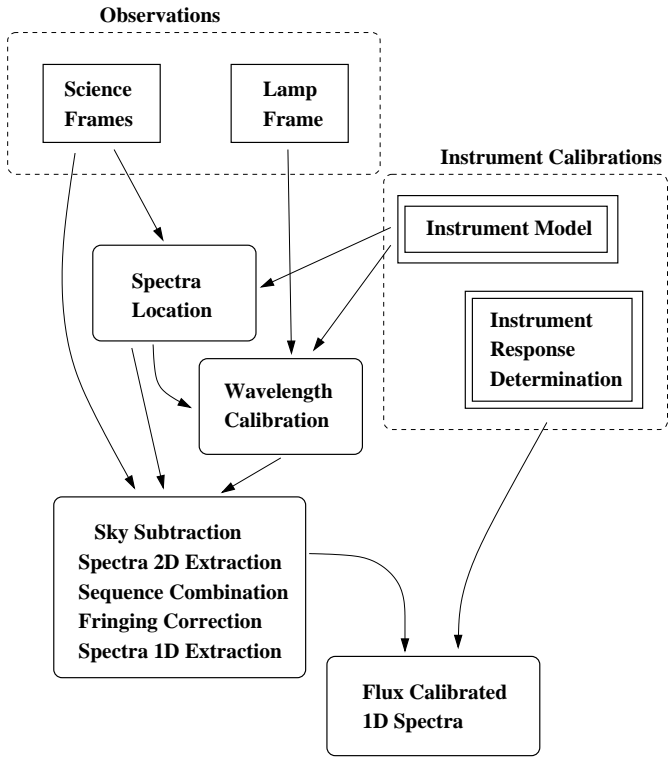}}
\caption{Block diagram summarizing the main steps involved in the
  reduction of MOS data using VIPGI.}
\label{fig:drs_scheme}
\end{figure}

The sky-subtracted slit spectra are then two-dimensionally extracted
using the tracing provided by the slit Curvature Model, and resampled
to a common linear wavelength scale. Only after this point the single
exposures of a jitter sequence are combined together. The main reason
to use a jitter sequence for the VVDS observations is to help
eliminate the strong fringing pattern present in VIMOS observations
obtained with the Low Resolution Red grism. All VVDS observations are
obtained using a 5 offsets pattern, with offsets of 0.75 arcsec
between exposures. For each slit, the two-dimensionally extracted
spectra from all exposures are median-combined a first time without
taking into account the jitter offsets, to produce a sky-subtraction
and fringing pattern residual map. Since the jitter offsets are
somewhat smaller that the typical object size, the median combination
is not enough to eliminate completely the object spectrum from these
residual maps, and therefore the object spectra areas are masked out
from the single slit spectra before the combination, interpolating
across adjacent sky-only parts of the spectrum. The residual map is
then subtracted from all single exposure spectra. At this point a
second combination is carried out, this time taking into account the
jitter offsets. The combination procedure uses the measured object
positions derived at the object detection stage to derive a very
accurate determination of the telescope offsets applied through the
jitter sequence. The single exposure residual-map-subtracted spectra
are offsetted to compensate for the effect of the jitter, and a final
average two-dimensional spectrum for each slit is
obtained. Figure~\ref{fig:fringing} shows two examples of such
two-dimensional spectra, as obtained with and without the subtraction
of the residuals map, to illustrate the effectiveness with which the
adopted procedure removes the fringing pattern from the data. This
removal is most efficient when only a faint object spectrum is present
in the slit, as in this case the adopted jitter pattern allows to
remove without problems the object spectrum from the no-offset
combination of the jittered exposures. As the objects get brighter and
larger their extended spectra can prevent a reasonable estimate of the
residual pattern to be obtained, and therfore the removal of the
fringing residuals becomes less accurate.


The object detection process is repeated on the combined spectra, to
produce the final catalog of detected spectra, and a one-dimensional
spectrum is extracted for each detected object using Horne optimal
extraction procedure (Horne, 1986). Finally spectra are flux
calibrated, using a simple polynomial fit to the instrument response
curve, derived from observations of spectrophotometric standard stars,
and corrected for telluric absorption features. The last correction is
based on a template absorption spectrum derived for each combined
jitter sequence from the data themselves.  Figure~\ref{fig:drs_scheme}
shows a block diagram summary of the various steps involved in the
reduction of a MOS observation with VIPGI.

\section{The quality of spectra extraction and wavelength calibration}
\label{sec:quality}

One rather obvious criterion we have used to assess the quality of our
data reduction pipeline is to compare its results with those obtained
via a purely manual data reduction carried out using
IRAF. Unfortunately the time requirements for a manual IRAF reduction
are rather expensive, and this comparison could be carried out only
for a few tens of spectra. It has shown nonetheless that the quality of
VIPGI-reduced spectra, in terms of continuum shape and signal to noise
ratio at all wavelengths, is basically the same of the one obtained
with a manual IRAF reduction. 

\begin{figure}
\resizebox{\hsize}{!}{\includegraphics{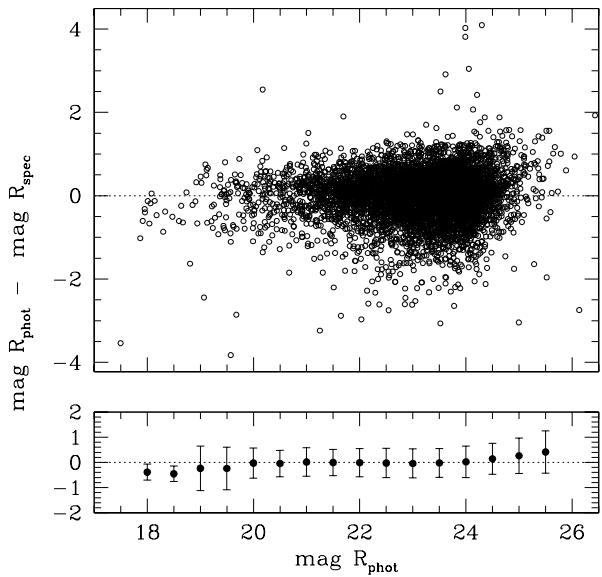}}
\caption{Comparison between the R-band magnitudes from the photometric
  VVDS catalog and the magnitudes computed integrating the observed
  spectrum with the equivalent filter response curve. The top panel
  shows the full distribution of data points, while the bottom panel
  shows mean and $1\sigma\ rms$ scatter in half magnitude bins.}
\label{fig:fluxes}
\end{figure}

More systematically, we have compared the R and I-band magnitudes from
the photometric catalog used to build the VVDS sample with the
magnitude derived integrating the flux in the observed spectrum with
the appropriate filter response curve. The results of such a
comparison are shown in Figure 5 of LeF{\` e}vre et al. (2004) for the
I-band magnitudes, and in Figure~\ref{fig:fluxes} for the R-band
ones. We can say there is a satisfactory agreement between the
spectral fluxes and the photometric magnitudes, although a significant
scatter can be seen in this comparison. Most of this scatter is
certainly due to the flux calibration procedure adopted for the
VVDS. In practice we are not trying to obtain a nightly flux
calibration, but we are using only a generic calibration template to
reproduce the correct spectral energy distribution shape, without any
specific attempt to obtain the right absolute flux normalization. Thus
variations in the sky transparency or in seeing conditions are not
taken into account by our flux calibration procedure, and show up in
the scatter visible in the figure.

The most important quality checks on VVDS data are the one about
wavelength calibration, and the accuracy and stability of redshift
measurements we can obtain with VVDS data reduced with VIPGI. As
discussed already in section~\ref{sec:lambdacal}, the accuracy of the
wavelength calibration relation measured on the lamp exposures does
not provide a complete picture of the global calibration accuracy for
VIMOS data.  We are therefore using the reduced spectra themselves to
assess this quality.


\begin{figure}
\resizebox{\hsize}{!}{\includegraphics{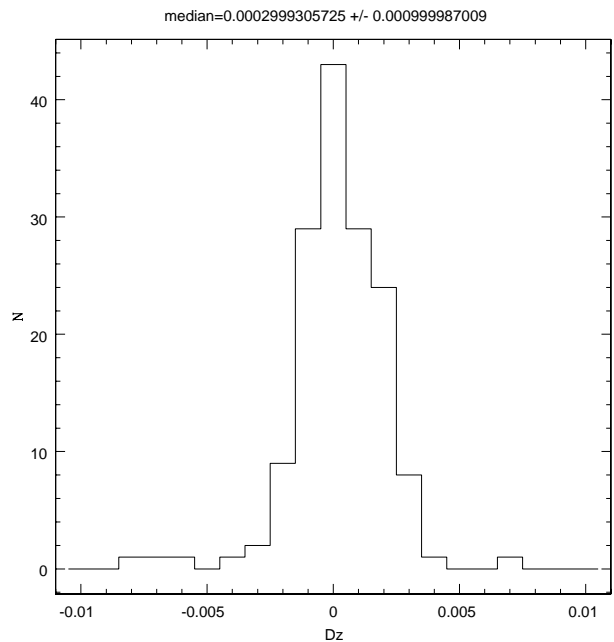}}
\caption{The distribution of differences in redshift measurements from
  repeated VVDS observations of a set of 150 objects.}
\label{fig:repeated}
\end{figure}

\begin{figure}
\resizebox{\hsize}{!}{\includegraphics{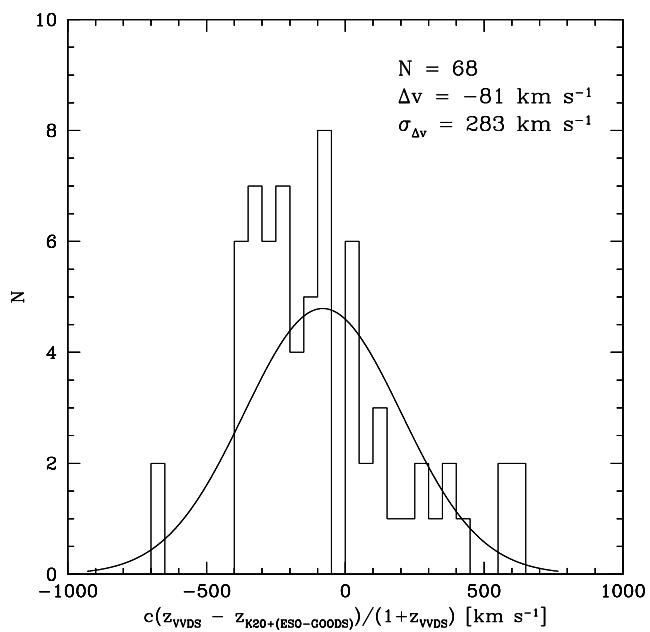}}
\caption{The distribution of differences in redshift measurements from
  a comparison between VVDS and K20 or GOODS observations.}
\label{fig:k20_goods}
\end{figure}

Using redshift measurements we can test the absolute stability of the
wavelength calibration, and this test can be carried out in two
different ways: within VIPGI, by comparing repeated observations of
the same objects, and via an external comparison, by comparing VIMOS
VVDS observations reduced with VIPGI with FORS2 observations obtained
by the K20 and GOODS surveys in the CDFS. As we are interested in
measuring the stability of the wavelength calibration, and not the
overall reliability of the redshift measurement process, we use in
this comparison only those spectra for which the separate redshift
measurements are in agreement (within a z interval of 0.1).  We have
150 repeated observations among the VVDS data which produce redshift
measurements in agreement between them: the differences have an
average value of 0.0003, with an rms scatter of 0.0010; the
distribution of these differences is shown in
Figure~\ref{fig:repeated}. We also have 41 and 27 galaxies in common
with the K20 and GOODS survey, with redshift measurements in
agreement. The differences have an average value of -0.0004 and
-0.0006, with an rms scatter of 0.0018 and 0.0013, respectively. The
global distribution of these differences is shown in
Figure~\ref{fig:k20_goods}.  Keeping into account the uncertainties
that are contributed to the redshift measurements by the
cross-correlation or line location procedures themselves, these
measured uncertainties in the redshift determinations allow us to say
that the absolute stability of the wavelength calibration in the VVDS
data reduced with VIPGI is at the level of 2 to 3 Angstrom,
significantly less than the linear dispersion coefficient for these
spectra (7.14 \AA/pixel).

One final measure of the quality of the spectra extraction process is
the fraction of failures in this process. There are many possible
sources for these failures, going from spurious detections in the
parent photometric catalog that result in no object being present in
the slits, or errors in the astrometric calibration of the catalog
resulting into a poor centering of the object in the slits, to slits
vignetted by the telescope guide camera probe, to problems in the sky
subtraction within a slit resulting into a failure in detecting a real
object spectrum, and to spectra characterized by an almost absent
continuum coupled with relatively strong emission lines that escape
detection. Overall, for the VVDS observations and the 16,936 slits
reduced thus far the total fraction of failures is approximately
3.9\%. Eliminating the cases where this failure is not due to the data
reduction software we can estimate that on average, if we exclude pure
emission line spectra objects, VIPGI is succesfully extracting a
spectrum for 98 \% of the slits where a real spectrum is present.

\section{Data Organization}
\label{sec:organizer}

As already discussed in Section~\ref{sec:VIPGI}, an automatic
reduction pipeline is not enough to guarantee a quick and timely data
reduction process for VIMOS data. An efficient organization of the
data and easy procedures for data browsing, to be used while assessing
the quality of the various data reduction steps, are two components as
important as the speed of the pipeline for the global data reduction
process. It is mainly for this reason that VIPGI provides its own,
VIMOS specific, data organizer.

\begin{figure*}
\resizebox{\hsize}{!}{\includegraphics{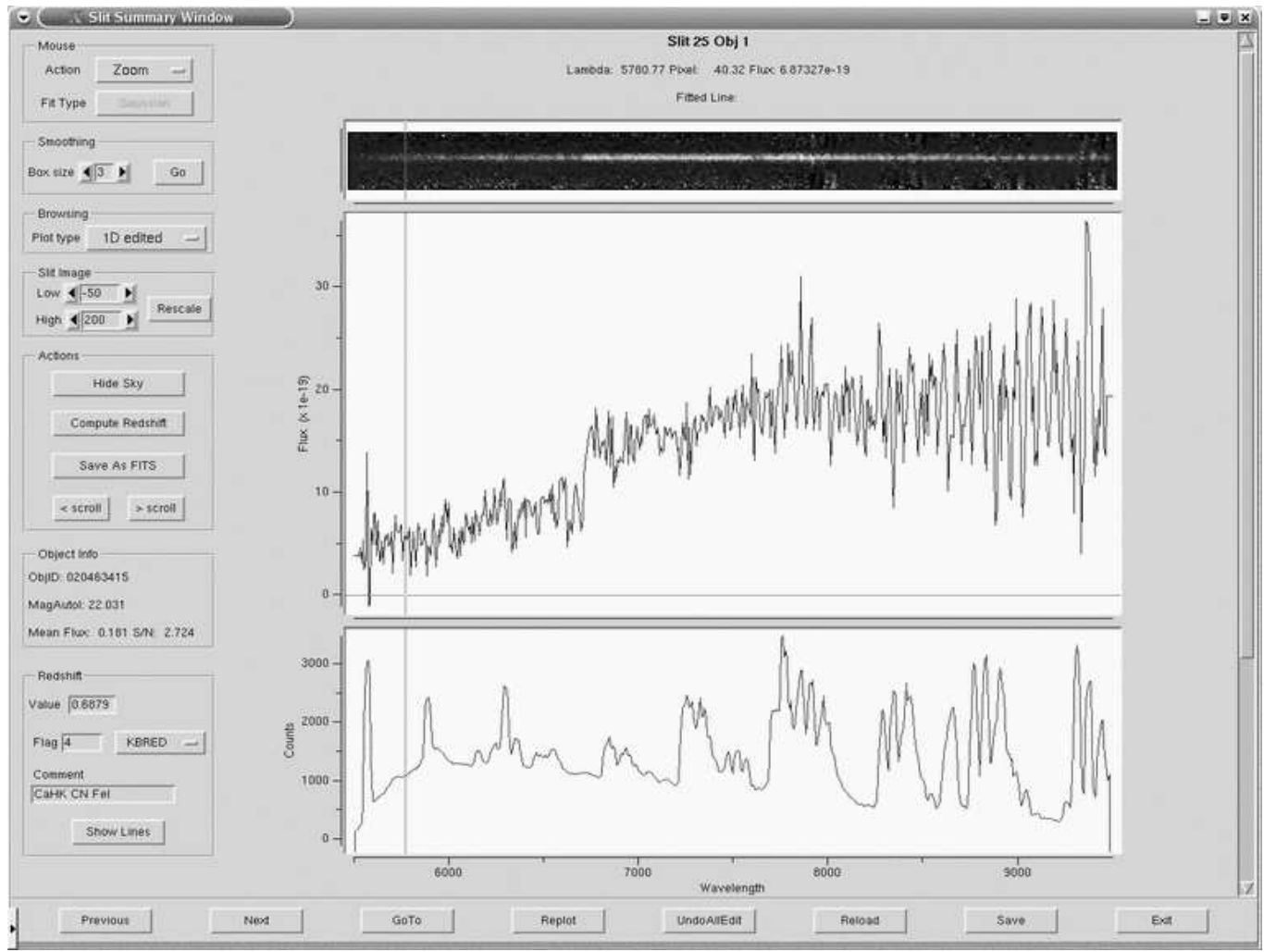}}
\caption{The ''slit summary plot'', showing the 2-dimensional and
  1-dimensional spectra for one VVDS object, and the sky spectrum
  for the same slit.}
\label{fig:summary}
\end{figure*}

VIMOS data are archived following standard ESO-VLT naming schemes, and
therefore the astronomer gets a set of FITS files named something like
VIMOS-2003-10-12T22:45:54.123.fits, names that do not convey any
information about the nature of the file they represent. Moreover
VIMOS observations typically result in a lot of files being produced,
at least four times those produced by a typical single CCD
instruments. To facilitate as much as possible the data reduction
process it was decided to build VIPGI around an automatic data
organizer that takes care of organizing the data in such a way as to
make mistakes while selecting input files for the various reduction
steps very unlikely to happen.

The first operation required in using VIPGI to reduce VIMOS data is to
import the data within the VIPGI data organizer structure; the raw
VIMOS data are copied from their location (a DVD from ESO archive or a
directory on the astronomer's computer disk) into a pre-defined
directory tree, divided by instrument mode (MOS, IFU, or Imaging) and
by quadrant.  Furthermore, they are also logically divided into
separate Data Categories and Types: a separate category for each MOS
mask, with arc lamp, flat field, and science exposures as separate
types; a separate category for each IFU target, plus one for arc
lamps, and one for flat field frames. Files are also renamed with a
naming scheme that reflects the logical categories used to group the
data.  As a result, it is very simple to define an homogeneous
data-set to be used as the input for any given data reduction
recipe. As part of the import operations, auxiliary calibration tables
needed by the pipeline recipes are also appended to the FITS files:
the ``CCD table'', containing informations about CCDs bad pixels; the
``GRISM table'', containing data about the spectra produced by the
grism used for the observation (length of the raw spectra, wavelength
interval over which to extract the 1-dimensional spectra, and over
which to carry out object detection before extraction); the ``LINE
CATALOG table'', containing wavelengths of arc lamp lines (or sky
lines in case one is using them for wavelength calibration); the
``Spectrophotometric table'', containing the flux measurements for a
spectrophotometric standard star.  The data organizer process
automatically appends to each file, according to its type, the
suitable tables.

\section{Spectra Plotting and Analysis}
\label{sec:plotting}

A number of different possibilities exist for browsing through
one-dimensionally and two-dimensionally extracted spectra.  The
simplest one is to display the reduced FITS file with an image display
tool like skycat, capable of handling in a simple way the display of
FITS image extensions.  In this way it is possible to browse quickly
through all stages of the data reduction process, as all the important
intermediate reduction products are appended as image or table
extensions to the output FITS file.\\
Besides the simple image display, ad-hoc display and plotting
functions have also been developed.  Two-dimensional slit spectra for
a MOS jitter sequence can be plotted together with all the
single-exposures 2D spectra used for the combination, to check the
reality of spectral features and the quality of fringing and sky
residuals removal.

A tool for plotting and analyzing the extracted one-dimensional
spectra is also provided. This tool allows the astronomer to plot each
one of the one-dimensional spectra, together with the corresponding
two-dimensional and sky 1D spectra.  In this way it is possible to
visually check the reality of spectral features that are present in
the one-dimensional spectrum, which could be due to sky, zero order or
fringing subtraction residuals.  It is also possible to zoom in and
out all three spectra, to edit the one-dimensional spectrum, smooth it
with a simple square window function, measure the signal to noise over
a selected wavelength interval, and fit the position of a selected
spectral line.  The astronomer can also obtain quick redshift
estimates by fitting or marking the position of a set of spectral
lines, and using a function that will compute a list of possible
redshifts based on a list of known emission and absorption lines in
galaxy spectra.  Once the user has chosen one possible solution, the
expected positions of all the lines in the list are marked on the plot,
to visually inspect the goodness of the redshift determination.

A ``summary'' plotting tool is also provided (see
Figure~\ref{fig:summary}), which incorporates in one display window
the functionality of the one-dimensional spectrum, of the lambda
calibration, and of the cross-dispersion slit profile plotting
tools. All functionalities from the one-dimensional plotting tool are
preserved. In addition, it is also possible to display information on
the astronomical object whose spectrum is being plotted.

\section{Summary}

We have developed VIPGI, a new tool designed to organize, reduce and
analyze data obtained with VIMOS, the imaging spectrograph built by
the VIRMOS Consortium for the European Southern Observatory.
This tool is being used to handle all the spectroscopic data obtained
for the VIMOS VLT Deep Survey, aiming to measure 150,000 galaxy
redshifts. 

VIPGI provides powerful data organizing capabilities, a small set of
data reduction recipes and dedicated data browsing and plotting tools
to check on the results of all critical data reduction steps, and to
plot and analyze final extracted 1D and 2D spectra.

We have performed many data quality check demonstrating that VIPGI can
be used to achieve a data reduction accuracy comparable to the one
obtained using standard IRAF tasks. Moreover, its high reduction speed
makes our pipeline the only reasonable tool to be used to reduce the
high amount of data that a modern instrument such as VIMOS produces.

VIPGI has been extensively used while reducing more than 20,000
spectra for the VIMOS VLT Deep Survey and it is now being used
routinely to reduce all kinds of VIMOS data in the VIMOS Data Reduction
Support Centers in Milano and Marseille.

\begin{acknowledgements}
We would like to thank Carlo Izzo, Ralf Palsa e Paola Sartoretti for
contributing to some parts of the code development, Tom Osterloo for
setting up the DRS coding infrastructure and the early code
development, and Nicolas Devillard for his many helpful suggestions.\\
This research has been developed within the framework of the VVDS
consortium.\\
This work has been partially supported by the
CNRS-INSU and its Programme National de Cosmologie (France),
and by Italian Ministry (MIUR) grants
COFIN2000 (MM02037133) and COFIN2003 (num.2003020150).\\
The VLT-VIMOS observations have been carried out on guaranteed
time (GTO) allocated by the European Southern Observatory (ESO)
to the VIRMOS consortium, under a contractual agreement between the
Centre National de la Recherche Scientifique of France, heading
a consortium of French and Italian institutes, and ESO,
to design, manufacture and test the VIMOS instrument.
\end{acknowledgements}

\end{document}